\newcommand {\nn} {{\nonumber}}
\newcommand {\fd} {\rightarrow}
\newcommand {\be} {\begin{equation}}
\newcommand {\ee} {\end  {equation}}
\newcommand {\ba} {\begin{eqnarray}}
\newcommand {\ea} {\end  {eqnarray}}
\newcommand {\bay} {\begin{array}}
\newcommand {\eay} {\end {array}}

\documentclass[12pt]{article}

\usepackage{graphicx}
\input english.sty

\begin{document}
\title{Critical behaviour and ultrametricity of 
Ising spin-glass with long-range interactions}
\author{Luca Leuzzi \\
{\small Dipartimento di Fisica, Universit\`a di Roma}
{\small {\em La Sapienza} }\\
{\small \ \  P. A. Moro 2, 00185 Roma (Italy)}\\}

\maketitle
\begin{abstract}
Ising spin-glass systems with long-range interactions ($J(r)\sim r^{-\sigma}$)
 are considered.
A numerical study of the critical behaviour is presented in the non-mean-field
region together with an analysis of the probability distribution of
the overlaps and of the ultrametric structure of the space of the equilibrium
 configurations in the frozen phase.  
Also in presence of diverging thermodynamical fluctuations 
at the critical point
the behaviour of the model  is shown  to be of the 
%-------\/
Replica Simmetry Breaking 
type and there are
 hints of a non-trivial ultrametric structure.
The parallel tempering algorithm has been used to simulate the  
dynamical approach to equilibrium of such systems.  
\end{abstract}

\section*{The Long-Range Spin-Glass Model}
The greatest incentive to study spin-glasses with long-range interactions is
that they are conceptually half-way between the 
%--------\/
Sherrington-Kirkpatrick (SK)
 model, exactly solvable
in mean-field theory, and the more realistic short-range models, with 
nearest-neighbour interactions. 
Long-range spin-glass models are particularly interesting because already in 
one dimension they show a phase transition between the paramagnetic and the 
spin-glass phase. So it is possible to study this transition, also out of the 
range of validity of mean-field approximation, in a 
%---------\/
relatively easy
 way 
in comparison with theories with short-range interactions below upper critical 
dimension.
Furthermore these one-dimensional models serve as  
a clarifying, qualitative, analogy 
for short-range models in higher dimensions.

\par The Hamiltonian of these kind of systems is:
\be
{\cal{H}}=-\sum_{i<k}J_{ik}s_is_k,
\label{lrham}
\ee
\noindent where $i=1,N$, the size of the system, the $s_i$ are Ising spin 
variables and the $J_{ik}$ are quenched, Gaussian 
random variables. They have mean zero and variance:
\be
\overline{J_{ik}^2}=\frac{C(\sigma)^2}{|i-k|^{2\sigma}},
\ee
\noindent where $C(\sigma)$ is just a normalizing factor, such that 
$\sum_{ik}\overline{J_{ik}^2}=N$; periodic boundary conditions have been used
(i.e. $i-k=N-i+k$ for $i-k>\frac{N}{2}$).

\par Already in   one dimension the long-range systems show different 
behaviours varying the value of $\sigma$. First of all, to allow 
thermodynamical convergence we must have $\sigma>1/2$ \cite{vave}. 
For $1/2<\sigma\leq 2/3$ a continuous phase transition is present, 
describable in the mean-field theory approximation;
the limit $\sigma_{mf}=2/3$ is found in the renormalization approach from 
the dimension of the coupling constant. Using the replica trick, in fact, 
we are able to get the Landau-Ginzburg effective 
Hamiltonian  corresponding to the Hamiltonian (\ref{lrham}). In $d$ dimensions 
it is:
\ba
&&{\cal{H}}=\frac{L^d}{4}\int \frac{d^dq}{(2\pi)^d}\left(
q^{2\sigma-d}+m^2_0\right)\sum_{a\neq b} 
\left|{\tilde{Q}}^{ab}(q)\right|^2 +
\label{form:lr_lg_ham}
\\
\nn
&&\: \: +\frac{g_0}{3!}
\int d^dx\sum_{a \neq b \neq c}Q^{ab}(x)Q^{bc}(x)Q^{ca}(x),
\ea
\noindent where $a$,$b$ and $c$ are the replica's indices
and the dimension of the coupling constant is, then, 
\be 
d_g=3\sigma-2 
\ee
\noindent for $d=1$. So $g$ is irrelevant for $\sigma<2/3$ 
(or marginal for $\sigma=2/3$).
When $2/3<\sigma<1$ the phase transition is supposed
%------\/
to  still be 
present but we are in an 
infrared divergent regime causing mean-field theory to lose consistency at the
critical temperature: 
it is then  necessary to renormalize in order to find the 
correct critical indices.
In the case $\sigma=1$ is not yet clear what kind of transition there is. 
Kotliar {\it{et al.}} \cite{kas} supposed a behaviour similar to the 
analogous case of the
long-range ordered magnetic systems with interactions decaying like $1/r^2$, 
in which Anderson {\it et al.} \cite{anderson} and then, 
in a version for general discrete models, Cardy \cite{cardy},
had found a 
%-----------\/
Kosterlitz-Thouless-like phase transition \cite{kt}.
Anyway nothing rigorous has been proved  until now for this value of $\sigma$.
Finally, for $\sigma>1$, there is only one Gibbs 
thermodynamical state at all temperature, as rigorously proved in
 \cite{c.o.vane.}.

There is, actually, an analogy  with the critical behaviour of 
short-range systems. 
Starting from  the lowest allowed value of the exponent driving the 
intensity of the bonds  and increasing it, we can observe 
%-------\/
behaviours 
qualitatively similar to those of short-range models in different dimensions:
from mean-field ($d\geq 6$) to infrared divergent regime and up to the case of
absence of phase transition.

\par Our contribution has been to determine the critical temperatures and the 
critical indices in the 
regime of diverging fluctuations at the critical point for different 
long-range systems (different values of $\sigma$).
Besides we have examined the features of the space of the equilibrium 
configurations for finite-volume systems, getting various hints about
the existence of a
%------------------------\/
 replica simmetry breaking (RSB) 
scenery also in the region of infrared divergences.
Furthermore, through the evolution in time of three independent replicas, 
we have got elements in favor of the existence of an ultrametric 
structure of the equilibrium states;
therefore sustaining the idea 
that this kind of phase space belongs  intrinsically to  spin-glasses, 
and that it does not depend on mean-field theory formulation, 
in which framework  it was initially derived.

\par We have done numerical simulations of these systems with different
 power-law behaviours, i.e. changing the value of the exponent $\sigma$. 
We took $\sigma=0.69$ and $\sigma=0.75$, both beyond $\sigma_{mf}$. 
They are the same values
chosen by Bhatt and Young in \cite{ba.yo.} so as to compare the 
common results.
Every system has been simulated in different sizes so to use the finite size 
scaling techniques: five  sizes between 32 and 512 spin have been 
investigated for every value of $\sigma$.
In this way we have got the critical indices and we have compared 
their values with
the theoretical values obtained from one-loop expansion in 
$\epsilon=\sigma-\sigma_{mf}$ \cite{kas}.
Moreover in every numerical run we have looked at the parallel evolution
of three independent replicas (observed in the same bond configuration).
In this way it has been possible to study observables built  from three 
different overlaps, that are useful to probe the ultrametric  structure
of the equilibrium configurations of a spin-glass, also out of the mean-field range of 
validity.

To simulate the dynamical approach to equilibrium we have used the parallel 
tempering algorithm \cite{giappo}. 
The evolution of every system 
has been simulated in a number of different random quenched samples varying 
between 200 (for $N=512$) and 600 ($N=32$), for 
65536 Monte Carlo steps each.
The thermalization times, in MC steps, were all at least 30 times smaller
 than this value. In order to find the  thermalization we have, first af all, 
checked  that the time that each configuration simulated in the parallel 
tempering spends in every heat bath is independent of the temperature of 
the bath. Also, we checked that the  probability of exchanging two 
configurations between two different baths is almost the same for every couple
of temperatures and it is always greater than $0.3$ for the set of 
temperatures chosen to perform the parallel tempering simulation.
Finally, the most important check on thermalization has been  to  
look at the absence of drifting  of the 
observables (the kurtosis of the overlap distribution and the 
spin glass susceptibility) on the logarithmic scale of time, after they 
reached their plateau values.

\section*{Critical Behaviour}

To determine the critical temperature we have used the 
finite size scaling (FSS) property of the observable:
\be
g=\frac{1}{2}\left(3-\frac{\overline{\left<q^4\right>}}
{\overline{(\left<q^2\right>)^2}}\right)
\ee
\noindent called {\it{Binder parameter}}. Here $< >$ stands for the mean 
over the thermodynamical {\it{ensemble}}, while the overline  represents 
the mean over the random distribution of the bonds. 
The overlap $q$ is defined, for our numerical 
goal, as:
\be
q=\sum_{i}s^{(1)}_is^{(2)}_i
\ee
where the upper index is the real replica's one.

The  finite size scaling form of the Binder parameter is:
\be
g={\overline{g}} \left( N^{\frac{1}{\nu}}(T-T_c) \right)
\label{gfss}
\ee
\noindent where $N$ is the size of the system. Since at $T=T_c$, 
for every size, is
$g=\overline{g}(0)$, the critical temperature can be deduced from different
sizes $g(T)$ intersections.
To compute it, we have used the scaling behaviour of the ``critical'' 
temperature for a finite size
 system:
\be
 T_c(N)-T_c^{\infty}=B\ N^{-\theta},
\label{tcfss}
\ee

\begin{figure}[!htb]
\begin{center}
   \begin{tabular}{rl}
      \includegraphics[width=0.49\textwidth,height = 0.4\textheight]{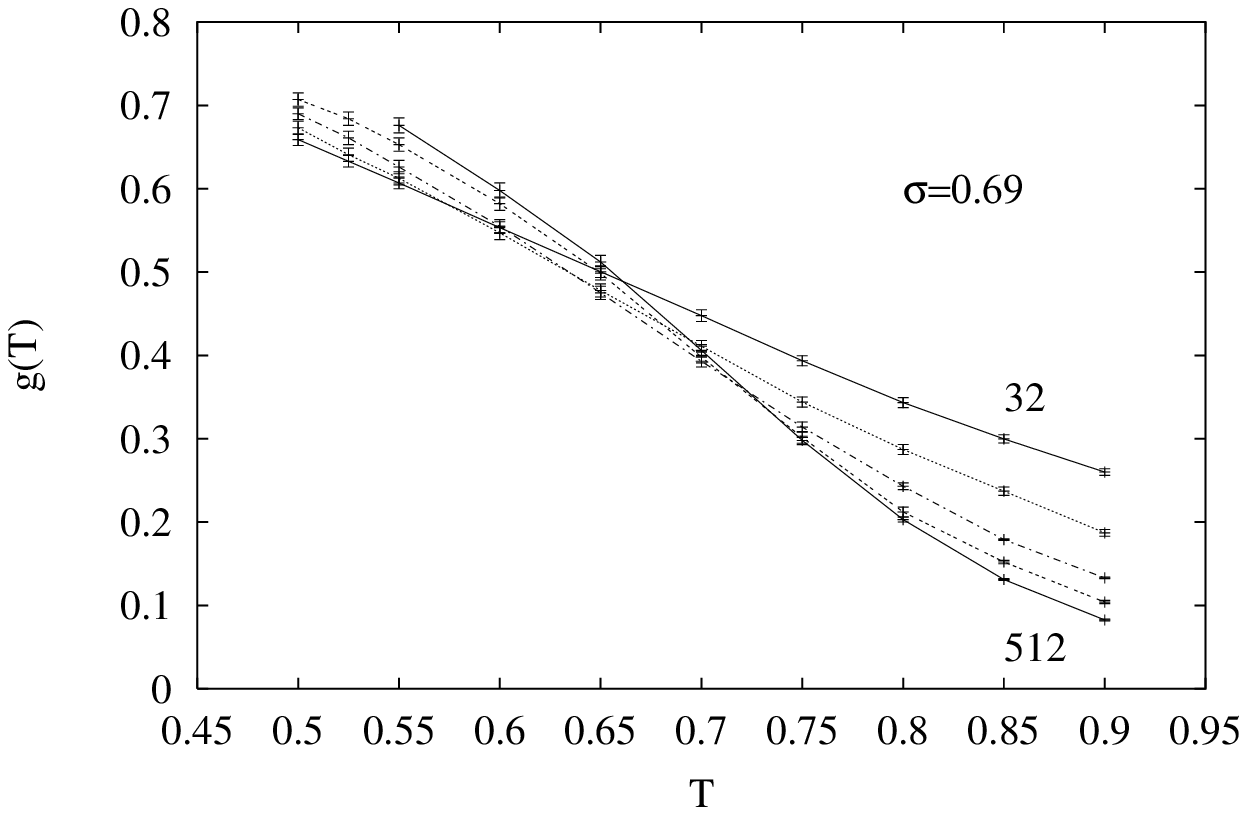}
      &
      \includegraphics[width=0.49\textwidth,height = 0.4\textheight]{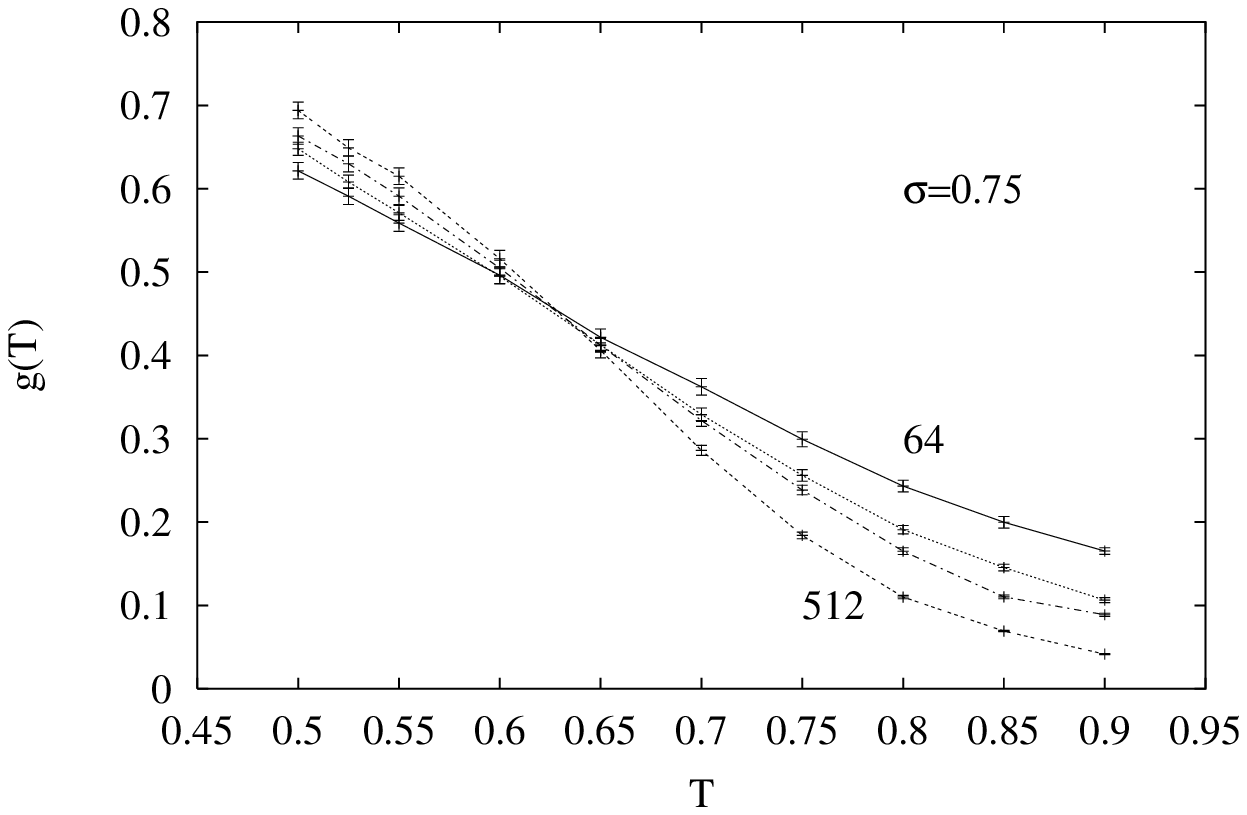}
   \end{tabular}
   \protect\caption{ Binder parameter $g$ vs. temperature for 
different sizes for the model with $\sigma=0.69$ (left) and the model 
with $\sigma=0.75$ (right). In the first case the 32 to 512 sizes are plotted,
 in the second one only the 64, 128, 256 and 512 are represented.}
   \label{fig:gBinder}
\end{center}
\end{figure}
{\par\noindent}where the $T_c(N)$ is
the abscissa of the intersection point between the $g(T)$ for the size $N/2$ 
and the $g(T)$ for the size $N$ and $\theta=1/\nu$ \cite{it-dr}.
Thus, fitting the curves of $T_c(N)$ for both 
long-range systems ($\sigma=0.69$ and $\sigma=0.75$) with a power-law
function, we have been able to extrapolate the following values for the 
critical temperature in the thermodynamical limit ($T_c^{\infty}$ in 
(\ref{tcfss})). 
We have got (see figure \ref{fig:gBinder}):
\ba
&&T_c=0.75\pm 0.1 \  \ , \mbox{for $\sigma=0.69$,} \\
&&T_c=0.63\pm 0.08 \ , \mbox{for $\sigma=0.75$.}
\label{form:tc}
\ea

\par
The first result is consistent with the two estimates of \cite{ba.yo.} 
for $\sigma=0.69$: $T_c \sim 0.73$ and
$T_c \sim 0.78$. However they couldn't localize the transition temperature
 for $\sigma=0.75$. Instead, we have found that
there is clearly a second order phase transition also at $\sigma=0.75$, well
besides the region of validity of the mean-field theory.
In analogy with short-range models, we observe that the behaviour of the 
Binder parameter is qualitatively similar to that of four-dimensional 
short-range spin-glasses, far below the upper critical dimension and 
 over the lower critical dimension (LCD) \cite{bhatt}
\cite{singh} \cite{parisi}.

\par
From the $g$'s FSS properties we have determined also the critical index 
$\nu$ \cite{iniparu}. We haven't used the value of 
the parameter $\theta=\frac{1}{\nu}$ computed from the fit  (\ref{tcfss}) 
because of its very large uncertainty.
Instead to estimate it we have first done the derivative of $g$ with respect 
to $T$ at a given value $g_0$. 
In fact, from (\ref{gfss}) follows:
\be
\left.\frac{dg}{dT}\right|_{T_0:g(T_0)=g_0}\simeq A L^{\frac{1}{\nu}}.
\ee

We have computed the values of the derivative for the values of $g$ 
corresponding to the confidence interval of the critical temperature.
In this interval of $g$-values we have fitted the $g(T)$ curves, 
in every size, with 
polynomials of various order (by the fact of second or third order),
 each time looking for the polynomial of the lowest possible order giving 
a  fit satisfying the $\chi^2$ test.

 For every $g_0$ value we have got different 
values of the $\left.\frac{dg}{dT}\right|_{T_0:g(T_0)=g_0}$ for 
different sizes. Then for every $g_0$ we determine a $\nu(g_0)$. 
The mean value of these
gives the correct exponent $\nu$.
\\
For our two models we have found:
\ba
\nu=3.8\pm 0.4          &,\mbox{  for $\sigma=0.69$}
\label{form:nu069}\\
\nu=4.5\pm 0.2          &,\mbox{  for $\sigma=0.75$}.
\label{form:nu075}
\ea
The first result is consistent with \cite{ba.yo.}, which gave $\nu=4.0\pm 0.8$,
and also with the  one-loop expansion result \cite{kas}: $\nu_{1l}=3+36\epsilon=3.84$
(here is $\epsilon=\sigma-2/3=0.69-2/3$).
For $\sigma=0.75$, instead, we are really too far  from $\sigma_{mf}$ for the 
one-loop expansion to give a good approximation ($\nu_{1l}=6$).

To find the critical index $\eta$ which gives the anomalous dimension of the 
two point correlation function at the critical temperature,
we have used the FSS properties of the 
observable $\chi_{sg}$, the so-called spin-glass susceptibility, defined as
\be
\chi_{sg}=\frac{1}{N}\sum_{ik}{\overline{\left(\left<s_is_k\right>\right)^2}}=
N{\overline{\left<q^2\right>}},
\ee
\noindent whose scaling behaviour is
\be
\chi_{sg}=N^{2-\eta}{\overline{\chi}}\left(N^{\frac{1}{\nu}}(T-T_c)\right).
\ee

We have got:
\ba
\eta=1.62\pm 0.08       &,\mbox{  for $\sigma=0.69$}\\
\eta=1.4 \pm 0.1        &,\mbox{  for $\sigma=0.75$}.
\ea
The theoretical value of $\eta$ in a long-range system 
, or, more exactly, its dependence on $\sigma$, such as that described
by the Hamiltonian (\ref{form:lr_lg_ham}), does not vary from the 
mean-field value going
in a region of diverging thermodynamical fluctuations, because the two-point
vertex function ($\Gamma^{(2)}$) does not have any infrared divergence at the 
critical point. 
The behaviour of the vertex function of the two {\it fields}
$Q^{ab}$ and $Q^{cd}$,
 as derived from the Hamiltonian (\ref{lrham}), is:
\\
\ba
\nn
\Gamma^{ab,cd}(k)&&=\left[\  k^{2\sigma-d}\right.-\  g_0^2(n-2)
\left.\int\frac{d^dp}{(2\pi)^d}\frac{1}{p^{2\sigma-d}\ (p-k)^{2\sigma-d}}
\right]F^{abcd}\equiv 
\\
&&\equiv \left[\  k^{2\sigma-d}-\  g_0^2(n-2)I_{\sigma}(k)\right]
F^{abcd}\equiv \\
\nn
&&\equiv \Gamma^{(2)}(k)F^{abcd},
\label{form:gamma2}
\ea
 where, in our case, $d=1$, $n$ is the number of the replicas 
and the tensor $F^{abcd}$ is defined like 
in \cite{green}:
\be
F^{abcd}=\frac{1}{2}\left(\delta^{ac}\delta^{bd}+\delta^{ad}\delta^{bc}-
T^{abcd}\right)
\ee
and 
\be
T^{abcd}=\left\{ \bay {ll} 1,&\mbox{ if $a=b=c=d$} \\
 0,&\mbox{ otherwise } \eay
\right.
\ee  
The integral can be easily computed:
\ba
\nn
I_{\sigma}(k)&&\equiv\int \frac{d^{d}p}{(2\pi)^{d}} \frac{1}{p^{2\sigma -d}} \frac{1}
{(p-k)^{2\sigma -d}}=\\
&&=\frac{(k^2)^{\frac{3d}{2}-2\sigma}}{(4\pi)^\frac{d}{2}} 
\frac{\Gamma\left(2\sigma -\frac{3}{2}d\right)}{\Gamma
\left(2d-2\sigma \right)}
\left[ \frac{\Gamma \left(d-\sigma \right)}{\Gamma \left(\sigma- \frac{d}{2} 
\right)} \right]^2.
\ea
Expressing the perturbative expansion in the variable
 $\epsilon=\sigma-\frac{2}{3}d$ we see that the function $I_{\sigma}(K)$ 
has no pole in $\epsilon$. Thus the  term $k^{2\sigma-d}$ in $\Gamma^{(2)}(k)$ 
of the free theory needs no correction from any perturbative contribution 
and the 
anomalous dimension $\eta$ of the two-points correlation function does not
depend on the order 
of the perturbation expansion. 

\begin{figure}[!htb]
\begin{center}
   \begin{tabular}{rl}
      \includegraphics[width=0.49\textwidth,height = 0.4\textheight]
	{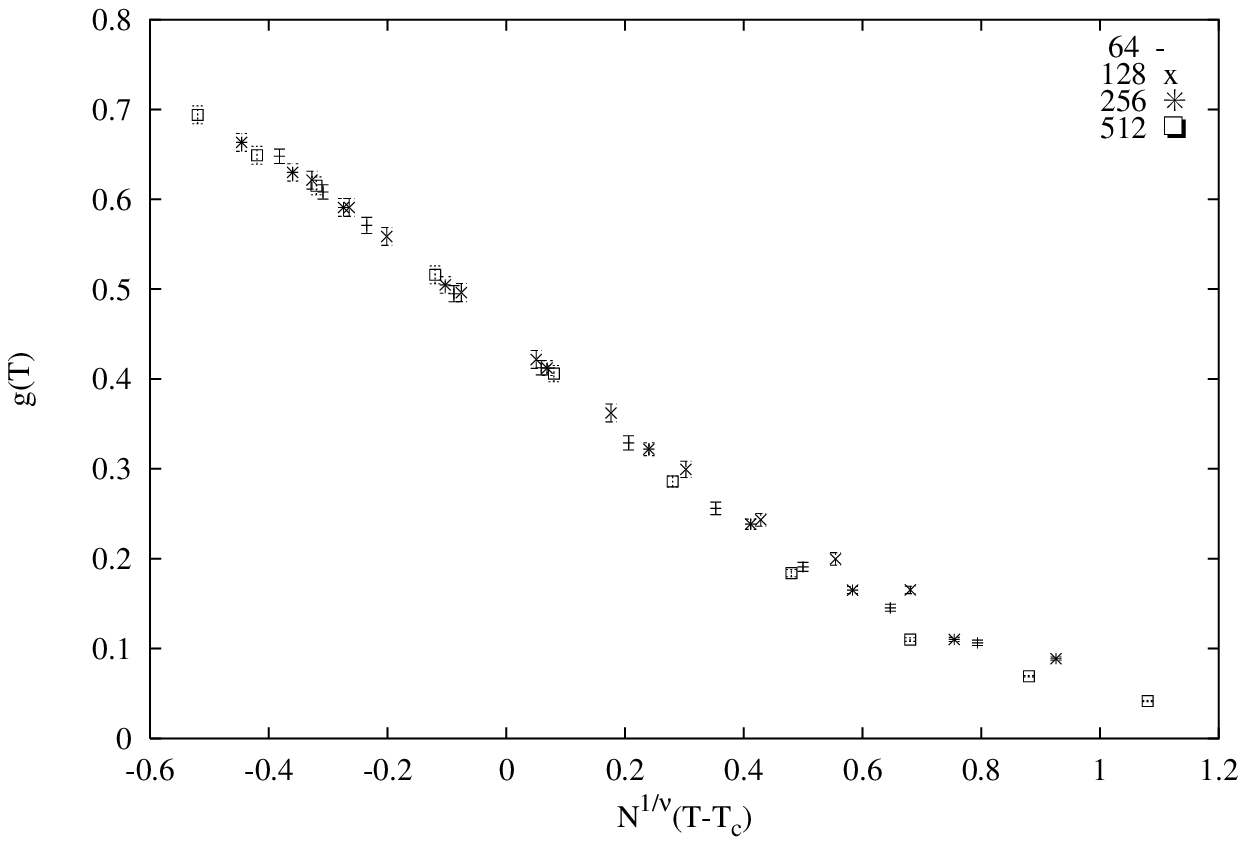}
      &
      \includegraphics[width=0.49\textwidth,height = 0.4\textheight]
	{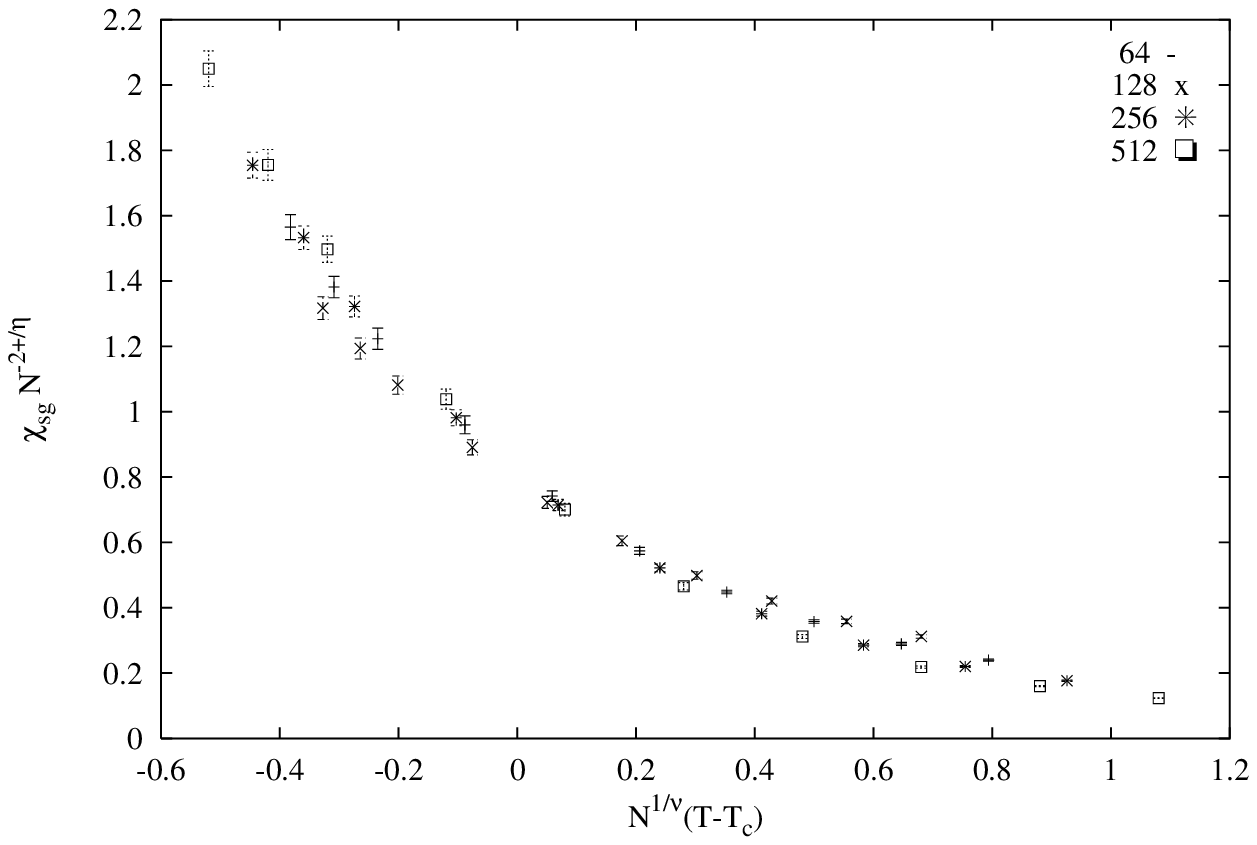}
   \end{tabular}
   \protect\caption{Scaling property of Binder parameter ({\it{left}}) and of 
spin-glass susceptibility ({\it{right}}). $g(T)$ and $\chi_{sg}N^{-2+\eta}$
 vs. $N^{\frac{1}{\nu}}(T-T_c)$ are plotted for all the sizes of the  
system with $\sigma=0.75$ ($T_c=0.63$, $\nu=4.5$, $\eta=1.4$, 
$N=64,128,256,512$).}
   \label{fig:scaling}
\end{center}
\end{figure}

{\par\noindent}Therefore, 
%------\/
it always has 
the same 
dependence on the exponent $\sigma$: 
 at any order is $\eta=d+2-2\sigma=3-2\sigma$. 
So the theoretical values are:

\ba
\eta_t=1.62       &,\mbox{  for $\sigma=0.69$}\\
\eta_t=1.5         &,\mbox{  for $\sigma=0.75$}
\ea
and  our results are in total agreement with them.
Using the so obtained values of the critical indices we can plot the scaling
behaviour of $g$ and of $\chi_{sg}$ as shown in figure \ref{fig:scaling}.

\section*{$P(q)$ Analysis and Ultrametricity}

The overlap probability distribution $P(q)$ is one of 
the most powerful means  at our
disposal to get 
information about the pure states structure of a spin-glass in its 
low temperature
 phase. 
The $P(q)$, in the standard replica approach, is connected to the 
structure of the {\it {finite-volume equilibrium states}}.
This point of view has been sometimes criticized, e.g. from Newman and Stein 
\cite{ns}, but some of the
objections raised by them have been overcome by showing that the
behaviour of the probability distribution functions built from {\it window} 
overlaps 
\footnote{The overlap over a window of linear size $B$ is defined as
\be
 q_B=\frac{1}{B^d}\sum^{d}_{i=1}\sum^{B-1}_{x_i}s^{(1)}(\{x_i\})
s^{(2)}(\{x_i\}),
\ee
 where $d$ is the dimension of the lattice, $x_i$ are the 
coordinates on the lattice and $B$ is smaller than $L$, the linear size 
of the lattice.}
is identical to that of  the functions $P(q)$ defined in the usual way 
\cite{small window}.

Here we present the data of the numerical simulations using the standard 
replica approach. However the data themselves do not depend on the definition 
of pure state: they are some kind of ``experimental'' facts
that must be explained by the theory.

\begin{figure}[!htb]
\begin{center}
   \begin{tabular}{rl}
      \includegraphics[width=0.49\textwidth,height = 0.4\textheight]{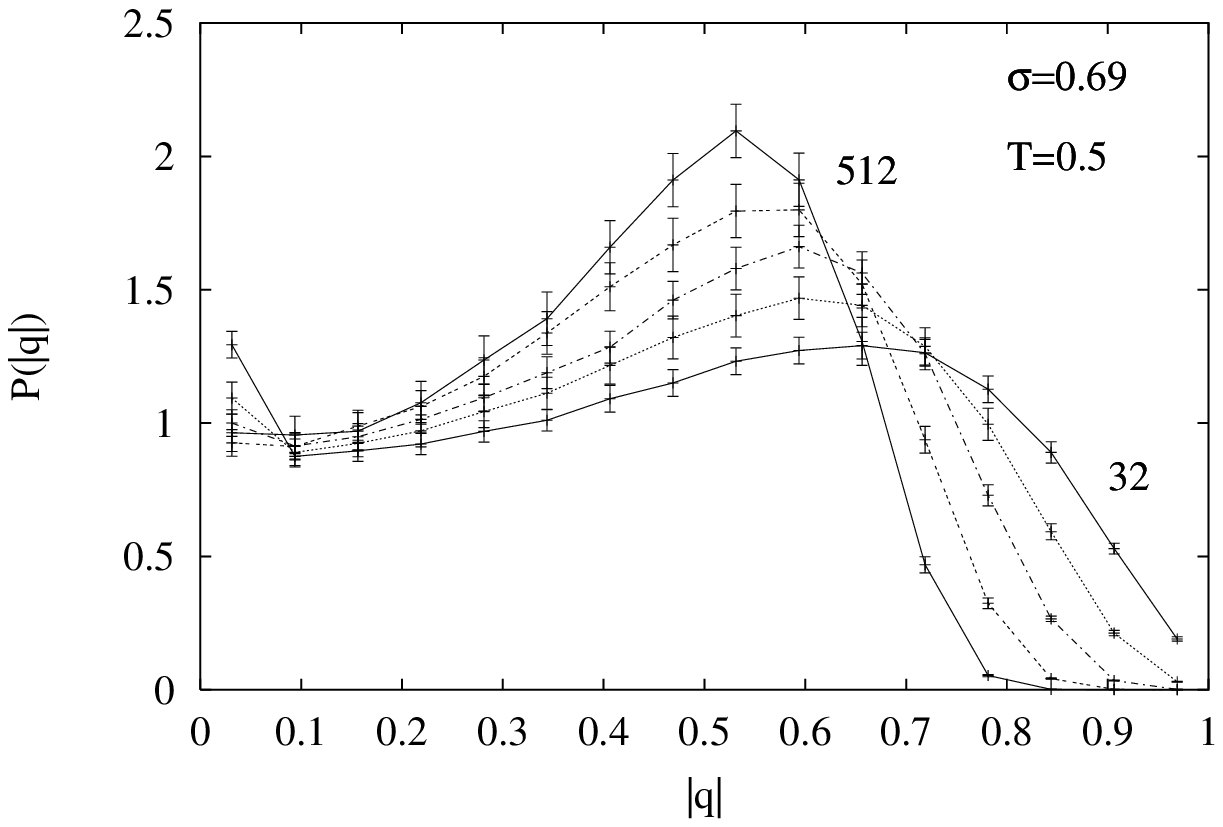}
      &
      \includegraphics[width=0.49\textwidth,height = 0.4\textheight]{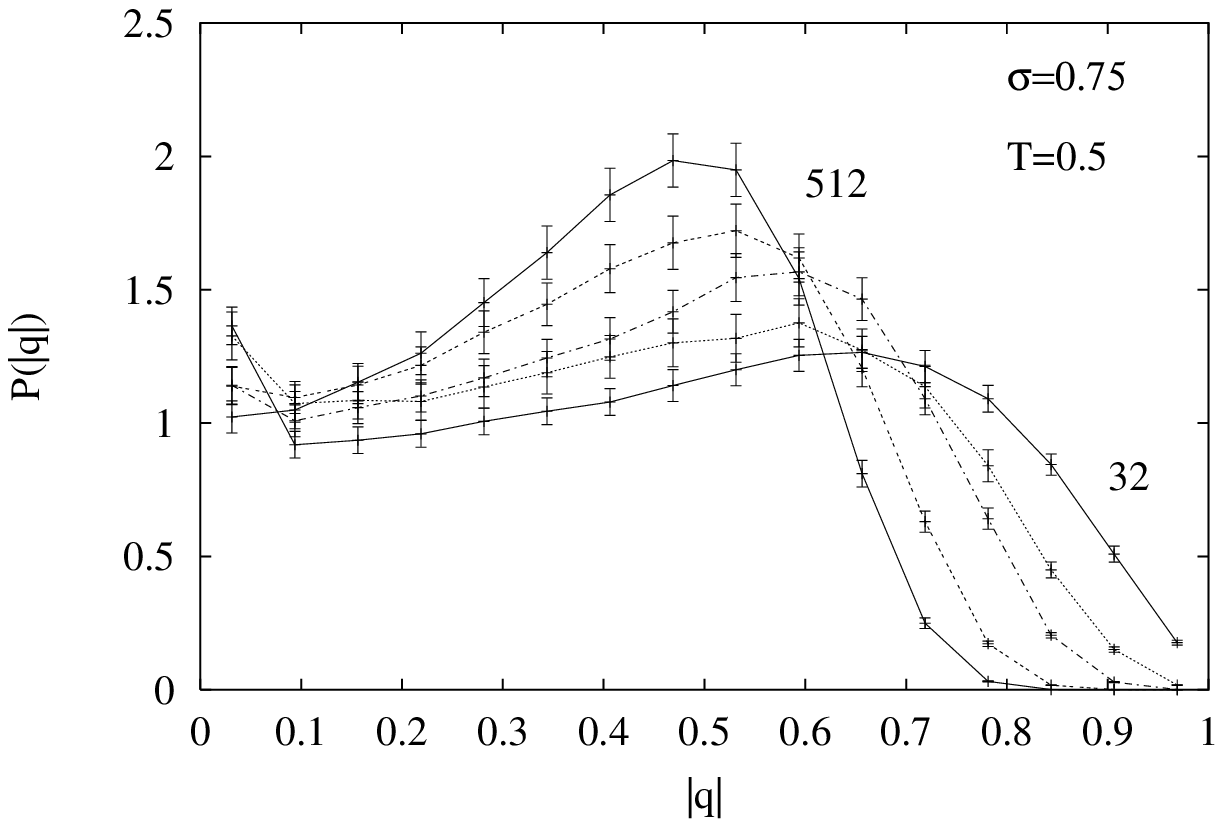}
   \end{tabular}
   \protect\caption{$P(q)$ at $T=0.5$ for the long-range models with 
$\sigma=0.69$ ({\it left}) and $\sigma=0.75$ ({\it right}). The sizes plotted 
are $N=32,64,128,256,512$.
Increasing the 
size the distribution flattens toward a value of about $1$, in the region 
$|q|\simeq 0$, and the peak become more and more sharp.}
   \label{fig:pdiq}
\end{center}
\end{figure}
An analysis of the behaviour of $P(q)$ allow us to discern between 
the RSB frame and the trivial one, in which only two different pure states
 are allowed.

Observing the behaviours of the probability distributions shown in figure
\ref{fig:pdiq},  we realize that we are  studying models, 
whose low temperature phase is described by  many equilibrium states, 
including those states which
are completely different and that corresponds to the region around $q\simeq 0$.
The peak  becomes higher as $N$ grows, while the area under the $P_N(q)$, 
between $q=0$ and the value of $q$ corresponding to the peak 
of the distribution, tends to remain constant. 
Furthermore $P(|q|=0)$ does not decrease, increasing the size of the system 
but 
%-------\/
settles down
 to a non zero value. 
This is the same picture we have in the 
mean-field case, 
%-\/--------------\/
also if we are now   considering systems which cannot be treated
in the mean-field approximation.
The fact that these distributions do not end with a $\delta$ function like the 
theoretical one in RSB theory,
but are non zero in the whole interval $[0,1]$, is an expected effect of 
  the finite size of the simulated systems.

\begin{figure}[!htb]
\begin{center}
   \begin{tabular}{rl}
      \includegraphics[width=0.49\textwidth,height = 0.4\textheight]
	{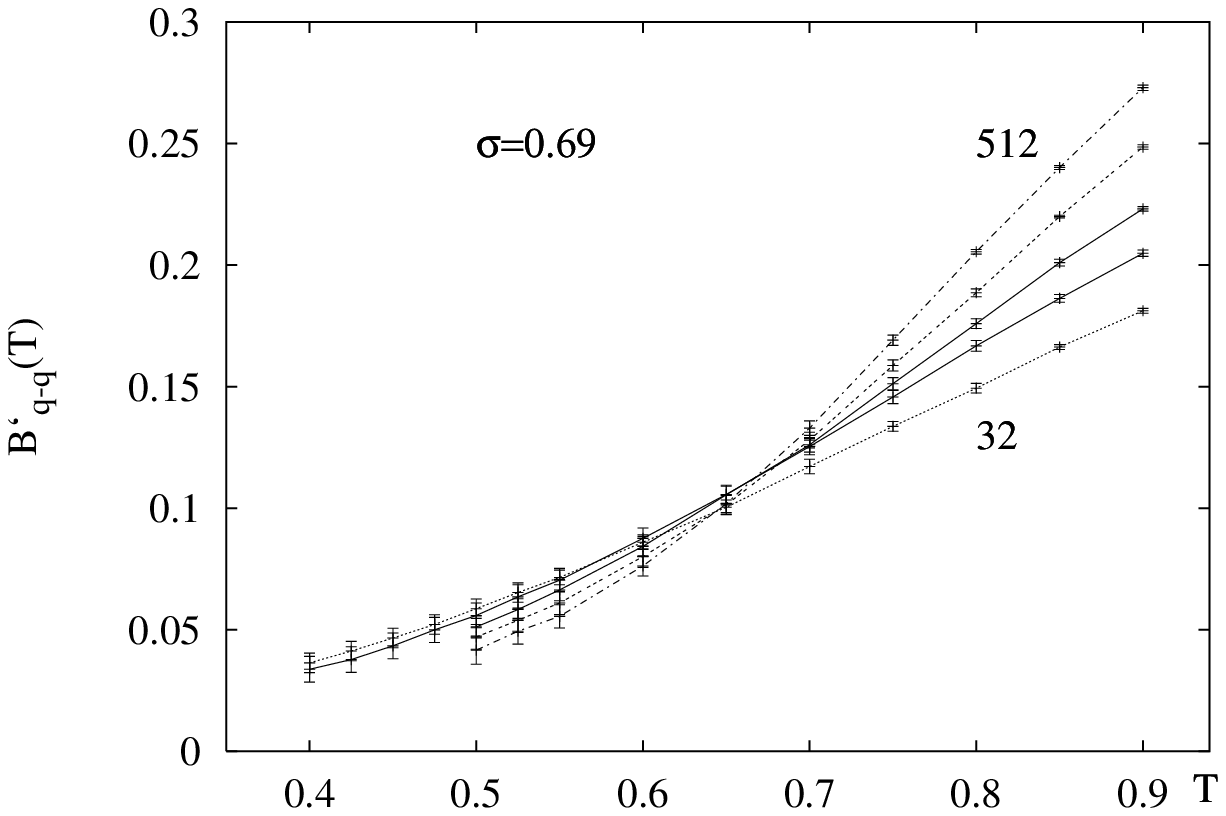}
      &
      \includegraphics[width=0.49\textwidth,height = 0.4\textheight]
	{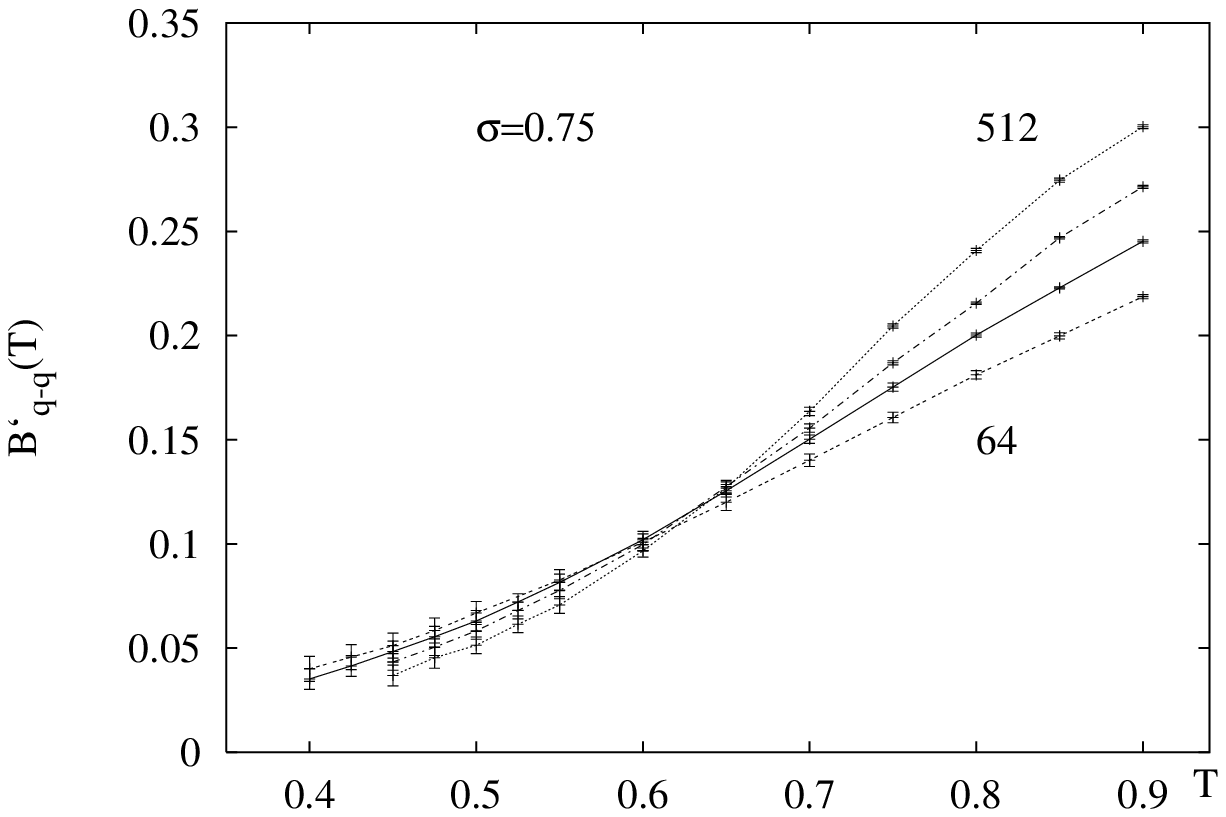}
   \end{tabular}
   \protect\caption{$B'_{q-q}(T)$ for the two long-range systems: 
	$\sigma=0.69$ ({\it{left}}), for sizes $N=32,64,128,256,512$, and 
	$\sigma=0.75$ ({\it{right}}), for sizes $N=64,128,256,512$.}
   \label{fig:bpq-q}
\end{center}
\end{figure}

\begin{figure}[!htb]
\begin{center}
  \begin{tabular}{rl}
      \includegraphics[width=0.49\textwidth,height = 0.4\textheight]
	{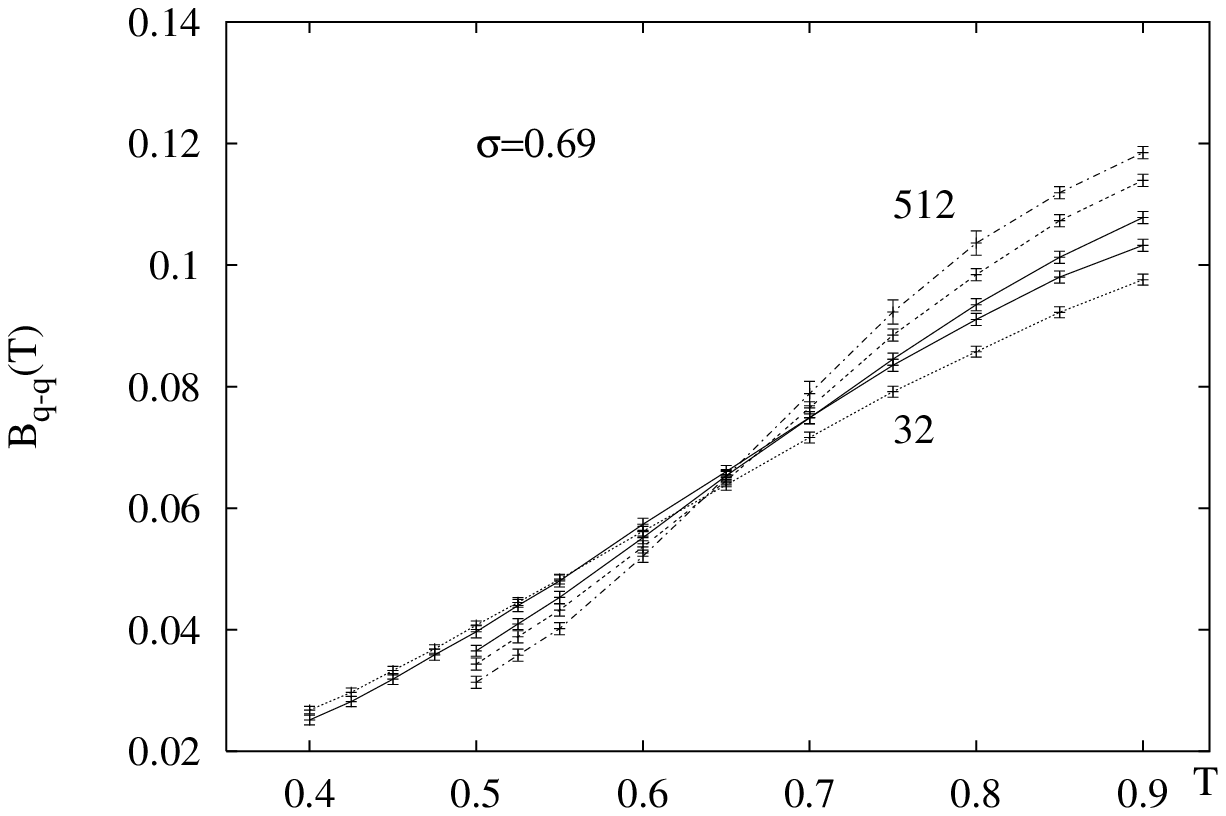}
      &
      \includegraphics[width=0.49\textwidth,height = 0.4\textheight]
	{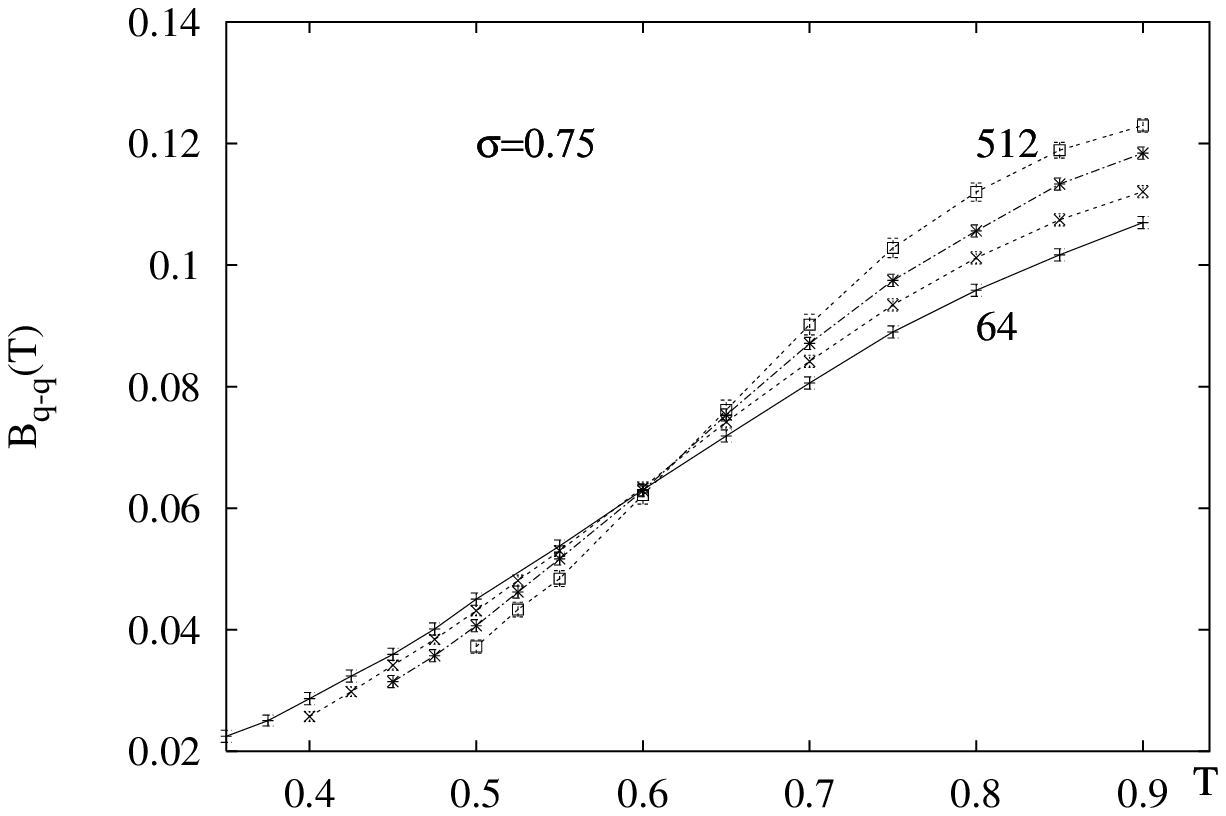}
   \end{tabular}
   \protect\caption{$B_{q-q}$ vs. temperature for the $\sigma=0.69$ model 
 and the $\sigma=0.75$ model ({\it{right}}). The sizes are the same as above.}
   \label{fig:bq-q}
\end{center}
\end{figure}

The most relevant and particular property  of the 
%-----\/
 phase space
at finite  volume
 that seems to emerge from our analysis is the ultrametricity, the special 
hierarchical structure of spin-glasses equilibrium configurations:
we can gather  hints of the existence of this property 
using some cumulants built from the overlaps $q_{12}$, $q_{13}$ and $q_{23}$
of three different, independent
replicas. 
With this aim we have observed the behaviour of the two cumulants
\ba
&&B_{q-q}\equiv\frac{{\overline{\left<\left(|q|-|q'|\right)^2\right>}}}
	{{\overline{\left<q^2_M\right>}}}  \ \ \mbox{ and } \\
&&B'_{q-q}\equiv
	\frac{{\overline{\left<\left(q-q'\mbox{sign}(q_M)\right)^2\right>}}}
	{{\overline{\left<q^2_M\right>}}}
\ea
where 
$q_M$ is the value of the overlap which has the maximum absolute value
 between the three, 
$q$ and $q'$ are the values of the other two overlaps ($q,q'<q_M$).
The  measures are made in every quenched configuration, at 
every temporal uncorrelated interval, once the simulated system has reached 
equilibrium.
%----------------------\/
Like the Binder parameter 
$g$, also these observables have a finite size scaling behaviour 
not depending on the index $\eta$. Their FSS form is, in fact,
\be
B^{\#}_{q-q}={\overline{f}}\left(N^{\frac{1}{\nu}}(T-T_c)\right).
\ee 
Analyzing their behaviours in the proper way we can get different checks
of the existence of a complicated space of states organized in an ultrametric
structure.

\begin{figure}[!htb]
\begin{center}
\begin{tabular}{rl}
\includegraphics[width=0.49\textwidth,height = 0.4\textheight]{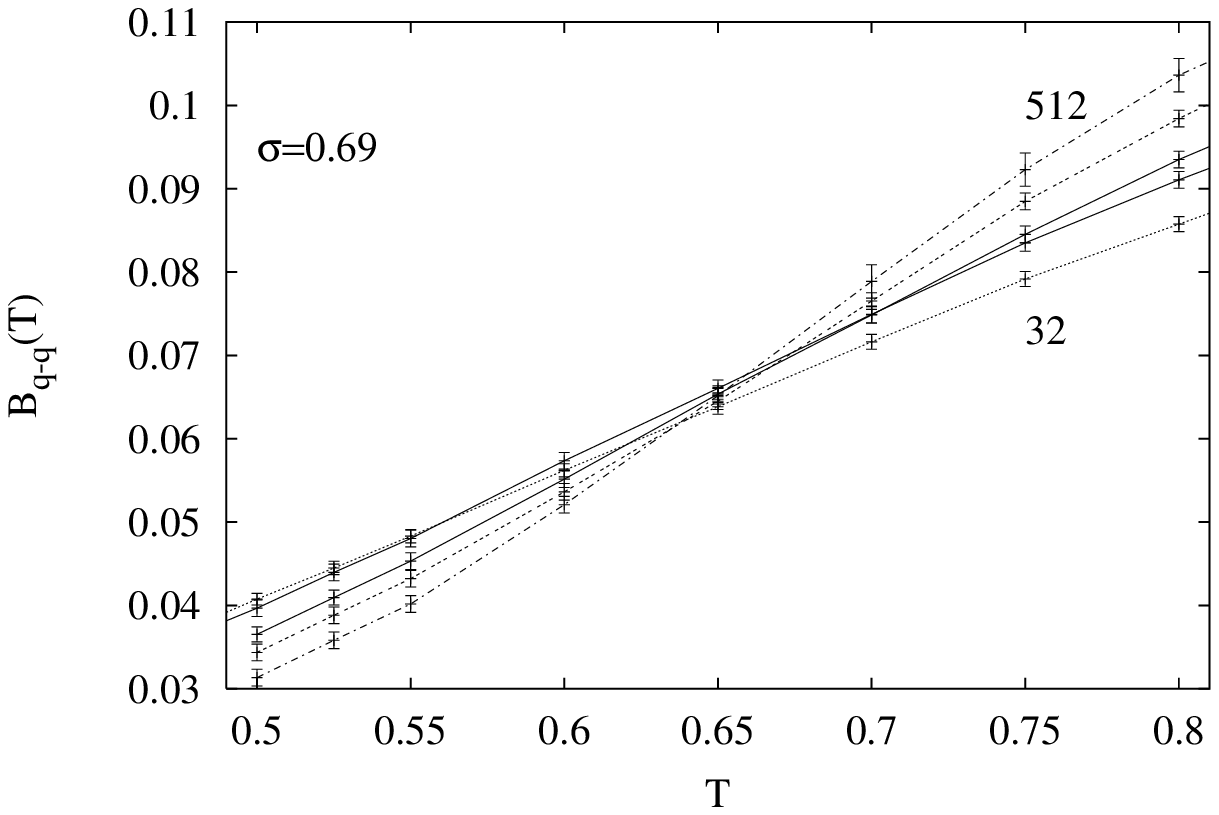}
    &
\includegraphics[width=0.49\textwidth,height = 0.4\textheight]{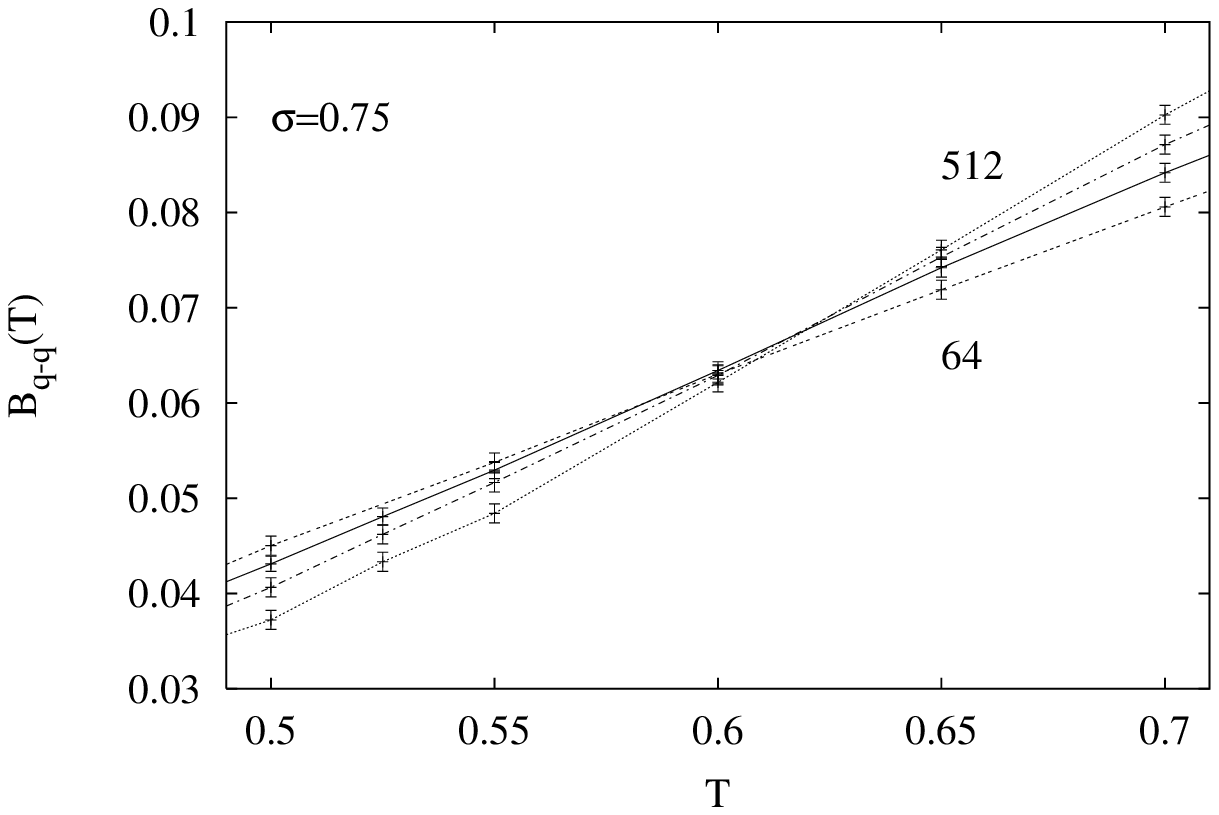}
\end{tabular}
\protect\caption{Detail of $B_{q-q}$ in the critical region
 for the $\sigma=0.69$({\it{left}}) and the
$\sigma=0.75$ model ({\it{right}}).}
\label{fig:bq-qtc}
\end{center}
\end{figure}

If an ultrametric structure exist the cumulants $B_{q-q}$ and $B_{q-q}'$ 
should go to zero under $T_c$ in the thermodynamical limit. The two minor
overlaps, in fact, should become equal and their difference should tend to 
zero.
This is, really, the behaviour that we  noticed and that is plotted in figures
\ref{fig:bpq-q} and \ref{fig:bq-q}:
 the quantities $B_{q-q}$ and the $B'_{q-q}$ of a 
given system  
%------------\/
tend to zero with decreasing   temperature and they do that
%-\/--------\/
the faster the  bigger is the size of the system.

At fixed  temperature below the critical one, $T=0.5$, we have fitted 
$B_{q-q}(N)$ with the power-law behaviour $A\ N^{-{\zeta}}$.
In both the long-range models considered, $B_{q-q}$,
at $T=0.5$, appear to decrease to zero  with this law.
The exponents are $\zeta=0.091 \pm 0.009$, for  $\sigma=0.69$, and
$\zeta=0.09 \pm 0.01$ for $\sigma=0.75$.
Because of these small values of $\zeta$, we would need data about
systems of greater size to guarantee that $B_{q-q}$ goes to zero below
the critical point.
Nevertheless the power-law decaying of the $B_{q-q}$ towards zero is consistent
with our data.

Following the behaviour from the high temperature phase the curves cross
each other in the critical region (we know it from the FSS behaviour)
and then tend  to zero for $T\fd 0$.
Actually, from this crossing we can have another guess of the critical 
temperature, just like from the Binder parameter $g$ (see figure 
\ref{fig:bq-qtc}).

In this case, however, there is no  fit of the FSS behaviour 
(\ref{tcfss}) satisfying the ${\chi}^2$ test. Thus we simply give the average 
of the last points of intersection between the $B_{q-q}(T)$ curves.  
The values found are:
\ba
&&T_c=0.65\pm 0.08 \ , \mbox{for $\sigma=0.69$}, \\
&&T_c=0.60\pm 0.06 \ , \mbox{for $\sigma=0.75$.}\nn
\label{form:tc2}
\ea
Anyway, this estimate agree, within the errors, 
with the previous one given in (\ref{form:tc}). The errors appear to be 
smaller than in (\ref{form:tc}), but we underline that 
we couldn't manage to do the FSS's fit, 
 neglecting, in this way, the
 shift of the $T_c(N)$ towards the $T_c$ of the system in the 
thermodynamical limit: the values above experience  a systematic error.

As we can note from the figures \ref{fig:bpq-q} and \ref{fig:bq-q}
the cumulant $B'_{q-q}$ is always greater
then $B_{q-q}$. This is due to the fact that not always $q'\mbox{sign}(q_M)$ 
has the same sign of $q$: there are triples of spin configurations giving 
products
$qq'q_M<0$, that is $\mbox{sign}(q'\mbox{sign}(q_M))\neq \mbox{sign}(q)$.
This implies that sometimes the contributions $(q-q'\mbox{sign}(q_M))^2$
 to the mean value are bigger
than the corresponding terms $(|q|-|q'|)^2$ in $B_{q-q}$. 
The qualitative behaviour of temperature
dependence, however, is not influenced in a critical way from this 
differences and 
$B'_{q-q}(T)$ goes to zero while $T\fd 0$ just like $B_{q-q}(T)$.
Fitting, as before, $B'_{q-q}$ at the fixed temperature $T=0.5$ with the 
power-law
$A'N^{{-\zeta}'}$, we observe a behaviour statistically consistent with 
the decaying to zero. The exponents are now ${\zeta}'=0.12 \pm 0.01$ for 
$\sigma=0.69$ and   ${\zeta}'=0.13 \pm 0.02$ for 
$\sigma=0.75$.

\section*{Conclusions}
In summary, the insight we get about the one-dimensional long-range 
($J(r) \sim \frac{1}{r^{\sigma}}$)
spin-glasses is that the critical behaviour satisfy the one-loop
predictions for $\sigma$ not too far from $\sigma_{mf}=\frac{2}{3}$, but that
the first-order
$\epsilon$-expansion fail to describe it already for $\sigma=0.75$.
In both examined systems we have been able 
to determine
 a low temperature phase, 
showing a non trivial hierarchical structure of the space of the  
finite-volume equilibrium configurations.
 We have built the $P(q)$ distribution, 
from which we can argue
 the validity of the RSB Ansatz also for $\sigma>\frac{2}{3}$,
out of mean-field theory, and we have analyzed the 
ultrametric structure of the equilibrium configurations
with the cumulants $B_{q-q}$ and $B'_{q-q}$.

To describe the behaviour of the one-dimensional long-range system with 
interactions decaying like $1/r$ ($\sigma=1$) a few analytical works have been 
made until now \cite{kas} \cite{moore}, based on the  replica symmetric
 Ansatz. 
Our results, however, show the 
inconsistency of this Ansatz in the explored region of parameters 
(i. e. until $\sigma=0.75$).
The Ansatz of replica symmetry is, furthermore, 
violated also in a related model 
for diluted infinite-range
systems \cite{bouchaud}.
Consequently further investigation should be done  to understand 
the behaviour of such power-law decaying systems for {$\sigma=1$}.

\medskip
{\bf {Acknowledgments}}
 I am deeply indebted with Professor Giorgio Parisi 
for his continuous support.

\end{document}